\documentclass[prl,twocolumn,showpacs]{revtex4} 
\usepackage{graphicx}% Include figure files
\usepackage{times}
\begin{document}
\title{Crumpling of a stiff tethered membrane}
\vspace{0.8cm}
\author {J.A. {\AA}str\"om$^1$, J. Timonen$^2$, and Mikko Karttunen$^3$} 
\affiliation{$^1$Centre for Scientific Computing,
P.O.\,Box 405, FIN-02101 Esbo, Finland}
\affiliation{$^2$Department of Physics, University of Jyv{\"a}skyl{\"a},
P.O.\,Box 35,           
FIN-40014 University of Jyv{\"a}skyl{\"a}, Finland}
\affiliation{$^3$Biophysics and Statistical Mechanics Group,
Laboratory of Computational Engineering,
Helsinki University of Technology,
P.O.\,Box 9203, FIN-02015 HUT, Finland}
\date {\today}

\begin{abstract}

A first-principles numerical simulation model for crumpling of a 
stiff tethered membrane is introduced. In our model membranes, 
wrinkles, ridge formation, ridge collapse, as well as the initiation of
stiffness divergence, are observed. The ratio of the amplitude and
wave length of the wrinkles, and the scaling exponent of the stiffness
divergence, are consistent with both theory and experiment. 
We observe that close to the stiffness
divergence there appears a crossover beyond which the elastic behavior
of a tethered membrane becomes similar to that of dry granular media.
This suggests that ridge formation in membranes and force-chain network
formation in granular packings are different manifestations of a single 
phenomenon.


%%% 46.25.-y Static elasticity
%%% 68.60.Bs Mechanical and acoustical properties
%%% 05.70.Np Interface and surface thermodynamics
%%% 82.20.Wt Computational modeling; simulation
\end{abstract}

\maketitle

Crumpling of, e.g., sheets of paper, is an everyday phenomenon that one easily 
passes without much extra attention. For physicists, however, crumpling has for a 
long time been a topic of intense research. In addition to fundamental statistical 
mechanic properties~\cite{Kantor:1986a},
issues such as scaling of strength and energy~\cite{t2,t3}, 
geometry and singularities~\cite{Cerda:1999a,t4}, and
even acoustic emission~\cite{Houle:1996a} and crumpling related 
phenomena in soft matter systems~\cite{Helfer:2001a},
have become topics of increasing interest.
In this article, we present a first-principles numerical simulation 
model for crumpling of a stiff tethered membrane, and relate
the problem to formation of force chain networks
in granular media~\cite{gn1,gn2}.

Crumpling of a thin elastic sheet or a stiff membrane demands deformation
energy. At very small strains a membrane is uniformly compressed.
Thin structures are, however, sensitive to buckling. At
first buckling appears as wrinkles. The wave length
and the amplitude of the wrinkles depend on the conditions of the
external loading and on the dimensions of the membrane~\cite{wrink}. 
As crumpling proceeds the deformation
energy begins to concentrate in narrow ridges and
conical peaks~\cite{t2,Cerda:1999a}. 
Upon crumpling of a piece of paper, for example,
the strain at the ridges and peaks become so large that irreversible
plastic yielding takes place. After the paper is stretched out
again, the ridge pattern can clearly be observed.

Another characteristic feature of membrane crumpling is that the effective
stiffness of a crumpled membrane increases fast when the degree of
crumpling becomes large. The ultimate limit of crumpling would be to press
a membrane into a more or less cubic form with a volume
corresponding to that of the volume of the membrane itself.
In practice this limit can hardly be reached. Squeezing a piece of
paper as hard as possible between the hands, results in a
ball that typically contains about 75\% air~\cite{mylar}. 
The behavior of the effective stiffness has been investigated by carefully
monitored experiments, and for  
a crumpled membrane it 
displays a power-law divergence as the dense packing limit is
approached~\cite{mylar}. The suggested explanation to the power-law divergence
is that the total ridge length diverges in a scale invariant way as
the porosity inside the crumpled membrane vanishes.

In order to study wrinkles, ridge formation, and the effective 
stiffness in crumpled membranes, we constructed a numerical 
model for stiff tethered membranes. The tethered membrane 
consists of frictionless spheres with stiffness $Y_s$ arranged 
in a triangular lattice. Each sphere has a mass $m$ and
moment of inertia $I$, and is connected to its neighbors 
by massless elastic beams which have both bending and tensile
stiffness.  It is reasonable to use
a higher Young's modulus for the spheres ($Y_s$) than for the beams ($Y_b$).
In the simulations here we chose $Y_s=10^7$ and 
$Y_b=5\times 10^3$. The membrane is placed
horizontally (i.e., in the $xy$ plane) in the middle of a rigid cube
of size $X^3(t)$, where $X$ is time dependent. Small random
fluctuations of the $z$ component of the centers of mass of the spheres
are introduced to avoid perfect symmetry. Newtonian dynamics is applied.

The environment inside the box is set to be strongly dissipative 
(the membrane can be thought of being immersed in a highly viscous fluid). 
This means that a force proportional to velocity, ($d\dot{\vec x}$), 
slows down the motion of each sphere. 
The membrane dimensions are considered to be large enough
for temperature fluctuations to be negligible (e.g. a paper sheet).
For computational reasons the membrane cannot contain more than about 
$10^5$ degrees of freedom, which sets a limit on its size. 
Wrinkles and the onset of ridge formation is nevertheless observed. 
Also the beginning of a power-law divergence of the effective
stiffness with an exponent close to the one observed
experimentally is seen. However, as porosity reaches a low enough
value, the effective stiffness begins to deviate from the
experimental power law. At low porosities the effective
stiffness of a tethered membranes behaves in the same way as
that of a packing of solid spheres~\cite{makse}. 

\begin{figure}[th]
\includegraphics[width=.4\textwidth]%,clip=true,viewport=0 5 680 485]
{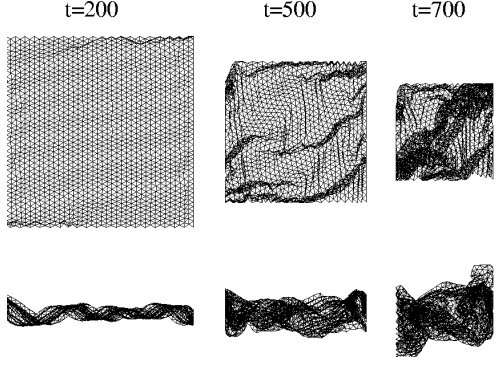}
\caption{Snapshots of a crumpling membrane. Top views (above) and side views
(below) at times $t=200,\,500,\,700$ are displayed. At $t=200$ wrinkles can be
observed. At $t=500$ ridges have formed and begun to collapse.
At $t=700$ the membrane is at the beginning of the stiffness scaling regime,
$(X-X_c)/X_c\approx 1.6$.}
\label{fig1}
\end{figure}

\begin{figure}[th]
\includegraphics[width=.38\textwidth]{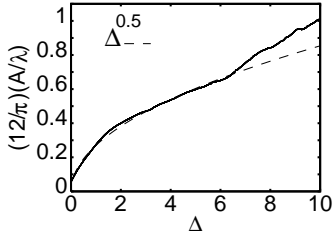}
\caption{
%%{\bf A} 
The ratio of wrinkle amplitude and wave-length
($A/\lambda$) as a function of membrane compression $\Delta$. 
The simulation result is compared to the theoretical prediction
$A/\lambda\propto\sqrt\Delta$.} 
\label{fig2}
\end{figure}

Figure~\ref{fig1} shows a snapshot from a simulation displaying a few wrinkles at $t=200$. 
By minimizing the elastic energy it can be demonstrated that
the relation between the wave-length $(\lambda)$ and the amplitude $(A)$
of the wrinkles is generally given by $A/\lambda\sim(\Delta/W)^{1/2}$, where
$W$ is the width and $\Delta$ the compression 
of the membrane perpendicular to the wrinkles~\cite{wrink}.
This relation is tested in Fig.~\ref{fig2} for the model membranes.
The square-root relation between $A/\lambda$ and
$\Delta$ holds up to $\Delta\approx 6$ (here $W\approx 23.4$). 

Deformation of ridges in controlled geometries
has been thoroughly investigated both
numerically and theoretically~\cite{t2,t1,t3,t4,t5,t6,t7,t8}.
The elastic deformation energy
of a ridge of length $L$ scales as $\kappa (L/\delta)^{1/3}$, where
$\delta$ is the thickness of the sheet and $\kappa$ its bending modulus.
By taking advantage of this result, and following the derivation in Ref.~\cite{mylar}, 
the scaling of stiffness of a membrane can be estimated: 
A crumpled sheet is divided into facets by the ridges.
Ridges surrounding a facet of size $L^2$ have a length proportional
to $L$. Such facets should fill a space of size
$L^3$. This leads to a bulk stiffness divergence of
the form $K=-VdP/dV\propto V^{-11/3}$, when the
volume $V$ surrounding the sheet decreases
under the influence of pressure $P$. This is not quite
consistent with experimental findings. In Ref.~\cite{mylar} $K$
was reported to scale as $K\propto (V_c - V)^{-\alpha}$
with $\alpha\approx 2.85$. Notice that this scaling relation
can also be expressed in the forms
$F\propto(\phi_c-\phi)^{-\alpha+1}$ and
$F\propto(X-X_c)^{-\alpha+1}$, where $F$ is the external
force applied on the membrane, $\phi$ the solid volume fraction,
and $X(t)$ the time dependent linear dimension of the crumpling
membrane.

\begin{figure}[th]
\includegraphics[clip=true,viewport=-6  90 81 180, width=.15\textwidth]{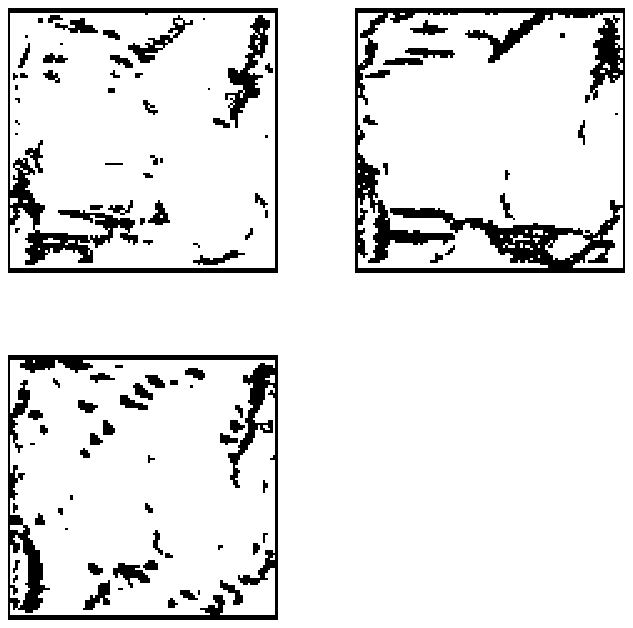}
\includegraphics[clip=true,viewport= 91 90 181 180, width=.15\textwidth]{af3.eps}
\includegraphics[clip=true,viewport=-6 -7 81 90, width=.15\textwidth]{af3.eps}\\
{\large \bf A} \hspace*{2cm} {\large \bf B}  \hspace*{2.2cm} {\large \bf C} \\
\caption{Contour maps (at $t=400$) of the spatial locations
of {\bf (A)} the highest local bendings $p(\vec x,t)<0.6$,
{\bf (B)} the highest elastic energy of the beams
$E_b(\vec x,t)$, and {\bf (C)} the highest contact
elastic energies of the spheres
$E_s(\vec x,t)$.}
\label{fig3}
\end{figure}

\begin{figure}[th]
\includegraphics[width=.38\textwidth]{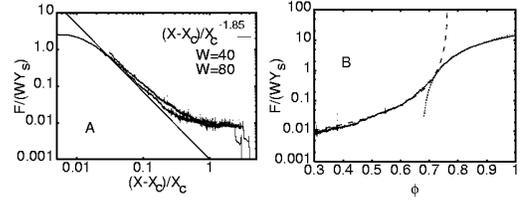}
\caption{{\bf (A)} The total force $F$ on the box
surrounding a membrane scaled by
sphere Young's modulus $Y_s$ and linear membrane length $W$
as function of $(X-X_c)/X_c$. Data is shown for membranes of linear size
$W=40,80$. 
The solid line shows the experimentally observed~\cite{mylar} power-law divergence
$F(X) \sim (X-X_c)^{-1.85}$. 
{\bf (B)} $F/(WY_s)$ as function of solid volume fraction $\phi$. 
The scaling of membrane stiffness $(\hat\phi_c-\phi)^{-1.85}$, with $\hat\phi_c\approx
0.76$, and the scaling of granular stiffness
$(\phi-\phi_c)^{1.62}$, with $\phi_c\approx 0.67$ are compared
to the simulation results.}
\label{fig4}
\end{figure}

To investigate possible ridge formation in the simulations, we
compare the local bending to the local energy.
We define the sum of the dot products of nearest neighbor bonds along the three
principal directions in a triangular lattice
as a measure of the local bending of the membrane,
\begin{equation}
p(\vec x,t)={1\over 3}\sum_{i=1}^3 {\vec l_i^n(t)\cdot\vec l_i^m(t)\over
l_i^nl_i^m}.
\label{eq:lbending}
\end{equation}
This is then compared to the local elastic energies of the beams
\begin{equation}
E_b(\vec x,t)=\sum_{i=1}^6\int_0^t {\bf F}(Y_b,\vec x(t))d\vec x(t).
\label{eq:lenergy}
\end{equation}
In Eq.~(\ref{eq:lbending}), $l^m$ and $l^n$ are the lengths of the 
beams on the opposite sides of a lattice site. In Eq.~(\ref{eq:lenergy}),
the elastic force ${\bf F}(Y_b,\vec x(t))$ is integrated
over the corresponding displacement and summed over
the six degrees of freedom for each site.

In addition, the elastic deformation energy of spheres in contact can be used to
identify ridges. It can be written as 
\begin{equation}
E_s(\vec x,t)=\sum_{i=1}^n {1\over 2}Y_s\delta_s^2,
\end{equation}
where $Y_s$ is the stiffness of the spheres and $\delta_s$ the virtual
'overlap' of spheres in contact. The sum is taken over all contacts.

Figure~\ref{fig3} shows contour plots of the lowest values of
$p(\vec x,t)$ and the highest values of
$E_b(\vec x,t)$ and $E_s(\vec x,t)$ at an early stage of
the crumpling process ($t=400$). It is quite obvious that 
$E_b$ and $E_s$ reach their highest values at the locations
of high local bending.
The patterns are quite complex, but elongated structures,
i.e. ridges, can be seen in the figure. As crumpling proceeds,
ridges grow and collapse and the membrane folds.
For later crumpling stages it is no longer possible to visibly detect
ridges as the patterns become too complex. 

The effective bulk stiffness of a membrane can be monitored, e.g.,
through its stress-strain curve. This is simply the total force
$F$ on the walls of the box surrounding the membrane, divided by
the linear size $W$ of the membrane ($F/W$) as a function of
$X(t)$. Initially $F(t)/W$ increases fairly linearly 
until the membrane begins to buckle and fold. As the dense packing
limit is approached, $F/W$ begins to increase
rapidly. Finally, when the membrane almost fills the box, a saturation
regime is entered. 

It is important to notice
that there are three types of dense packing limits:
1) The tightest packing is achieved when the membrane volume 
is equal to the volume of its container. In this limit, the solid volume 
fraction is $\phi_1=1.0$. 2) 
Hard-core spheres. Here,
the tightest packing is an FCC lattice with
$\phi_2= \pi\sqrt{2}/6\approx 0.74$. 
3) A random dense packing of spheres
with $\phi_3\approx 0.63$.
For membranes it seems that the stiffness divergence takes place 
at $\phi_c\approx 0.75$ which is close to $\phi_2$.
The regime of saturating stiffness is, however, entered well before
this limit is reached. 

Figure~\ref{fig4}A shows $F/(Y_sW)$ for membranes with $10^4$ and
$4\times 10^4$ degrees of freedom (each simulation demands about
$10^6-10^7$ time steps). The simulation results are compared to the 
experimental result reported by Matan {\it et al.}~\cite{mylar},
who found that $F(X)\propto (X-X_c)^{-\alpha+1}$, with
$\alpha\approx 2.85$. Figure~\ref{fig4}A shows good agreement
between simulations and experiment until saturation
sets in at $(X-X_c)/X_c\approx 0.03$. 

At saturation there is a crossover to a different type of stiffness behavior.
Beyond saturation the beams no longer affect the stiffness of the membranes.
Instead, the membranes begin to behave like granular packings~\cite{makse}.
The stiffness of a 3D packing of solid
spherical grains scales like $(\phi-\phi_c)^\beta$,
with $\beta\approx 1.62$, and the critical volume is
$\phi_c\approx 0.6-0.7$ depending on the form of
interaction between the grains.
Figure~\ref{fig4}B shows a comparison between simulations and the scaling
behavior of $F(\phi)$ in the two mentioned regimes.

\begin{figure}[th]
\includegraphics[width=.38\textwidth]{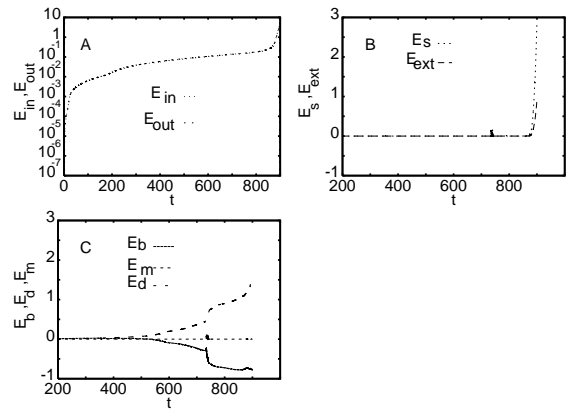}
\caption{{\bf (A)} Internal and external energies as functions of time
$t$. {\bf (B)} $E_s$ and $E_{ext}$ as functions of $t$. {\bf (C)}
$E_d$, $E_b$ and $E_m$ as functions of $t$. }
\label{fig5}
\end{figure}

By following the energy contents in a membrane from the initially 
flat membrane to a fully crumpled state, it is possible to 
determine the energy component responsible for the stiffness 
divergence seen in Fig.~\ref{fig4}. The external energy 
$\int_t F(t)\left [ X(t)-X(0)\right ]dt$ is compared to the 
internal energy of the membrane in Fig.~\ref{fig5}A. 
There is no difference between the  
external and internal energies as energy conservation also demands.

The internal energy can be split into five terms that are related
to the force terms in the general equation of motion
\begin{equation}
m\ddot{\vec x} + d\dot{\vec x} + {\bf K}{\vec x} +
Y_s\vec\delta_s =\vec f_{ext}.
\end{equation}
Here, $x$ is the displacement, $m$ the mass, $d$ the damping coefficient,
${\bf K}$ the stiffness matrix, $\vec\delta_s$ the
deformations (or virtual overlap) of spheres in contact, and $\vec f_{ext}$
are forces applied on the walls of the box surrounding the membrane.
The related energy terms are the following:
${\bf (1)}$ The elastic energy of the beams, $E_b$. 
${\bf (2)}$ The compression energy at the contacts
between spheres, $E_s$. 
${\bf (3)}$ The compression energy of the spheres against the external
walls, $E_{ext}=0.5\sum Y_w\delta_w^2$. 
${\bf (4)}$ The energy dissipated through damping,
$E_d=\int_t \vec F_dd\vec x$. 
${\bf (5)}$ The kinetic energy of the spheres, $E_m=0.5\sum m\dot{\vec x}^2$. 
These energy
components are shown separately as functions of time in Figs.~\ref{fig5}B and
\ref{fig5}C. The diverging terms are $E_s$ and $E_{ext}$. The divergence of
$E_{ext}$ is rather trivial as it is just the deformation energy
of the box surrounding the membrane. It is thus the energy
$E_s$ that causes the stiffness divergence. This explains to some
extent why there is a crossover to the stiffness behavior of
granular packings. When the deformation energy at the contacts
between spheres dominate over other energy terms, the
beams no longer play a role in the membrane stiffness. 

\begin{figure}[th]
\includegraphics[width=.38\textwidth]{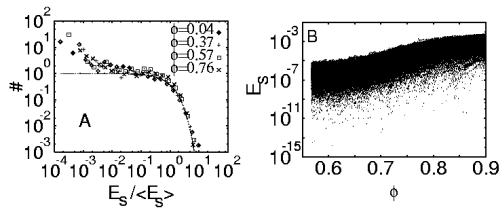}
\caption{{\bf (A)} Distribution functions
for the deformation energies at contacts between spheres for different solid
volume fractions $\phi=0.04,0.37,0.57,0.76$. The distribution functions
are compared to the equilibrium distribution $\exp(-E_s)$~\cite{OH}.
{\bf (A)} $E_s$ as a function of $\phi$. The crossover that takes place
at $\phi\approx 0.72$ is not visible in the data.}
\label{fig6}
\end{figure}

Since there are two different scaling behaviors of stiffness,
and both regimes are dominated by $E_s$, one would expect that at the 
crossover point there appears a qualitative change in the 
function describing the distribution of contact energies
of the spheres. There is, however, no indication of such a qualitative change
as is demonstrated in Fig.~\ref{fig6}, which means that, at the crossover,
ridges are simultaneously force chains. When comparing the topology
of the network of ridges in crumpled membranes and that of the force-chain 
network in dry granular materials~\cite{gn1,gn2}, one notices that both are
essentially one-dimensional lines of localized deformation
energy, and form complicated networks in three-dimensional space. 
Based on these observations, it is plausible to draw
the conclusion that the ridge formation and force-chain formation
are not separate phenomena, but different manifestations of a single 
phenomenon. The crossover between
the two then only marks the region where the membrane thickness
begins to limit the density of network lines.
  
In conclusion, we have demonstrated for the first time 
that a numerical model of a tethered membrane can reproduce the theoretically
predicted wrinkling and ridge formation, and experimentally observed 
stiffness divergence. The early-stage ridges can be traced both by the 
elastic deformations of the beams and by the contact deformations of
spheres in a tethered membrane. The divergence of
the effective stiffness is a result of the divergence of the contact 
energy of the spheres. This eventually leads to a crossover
in the stiffness between two different scaling behaviors. When the
solid volume fraction is far below that of the random dense 
limit of granular packings, the 
scaling of stiffness is 'membranic', and when the random dense packing
is approached, the stiffness scaling becomes 'granular'. There is no
obvious difference in the distribution of deformation energy
in the two regimes. It thus seems that deformation ridges in
membranes and force chains in granular packings are different 
manifestations of the same phenomenon.

\end{document}